\RequirePackage{ifpdf}
\ifpdf 
\documentclass[pdftex]{sigma}
\else
\documentclass{sigma}
\fi

\begin{document}
\allowdisplaybreaks

\renewcommand{\PaperNumber}{011}

\FirstPageHeading

\ShortArticleName{Order Parameters in XXZ-Type Spin $\frac{1}{2}$ Quantum Models with Gibbsian Ground  States}

\ArticleName{Order Parameters in XXZ-Type Spin $\boldsymbol{\frac{1}{2}}$\\
Quantum Models with Gibbsian Ground  States}

\Author{Wolodymyr SKRYPNIK}

\AuthorNameForHeading{W.~Skrypnik}

\Address{Institute of Mathematics, 3 Tereshchenkivs'ka Str., Kyiv
4, 01601 Ukraine}
\Email{\href{mailto:skrypnik@imath.kiev.ua}{skrypnik@imath.kiev.ua}}

\ArticleDates{Received October 19, 2005, in final form January 16,
2006; Published online January 24, 2006}

\Abstract{A class of general spin $\frac{1}{2}$ lattice models on
hyper-cubic lattice $Z^d$, whose Hamiltonians are sums of two
functions depending on the Pauli matrices $S^1$, $S^2$ and $S^3$,
respectively, are found, which have Gibbsian eigen (ground) states
and two order parameters for two spin components $x$, $z$
simultaneously for large values of the parameter $\alpha$ playing
the role of the inverse temperature. It is shown that the
ferromagnetic order in $x$ direction exists for all dimensions
$d\geq 1$ for a wide class of considered models (a proof is
remarkably simple).}


\Classification{82B10; 82B20; 82B26}

\section{Introduction and main result}

The existence of several long-range orders (lro's) and order
parameters in quantum many-body systems is an important problem
which is the first step towards a description of their phase
diagrams.

In our previous paper \cite{Dorlas&Skrypnik} we found a class of
quantum spin $\frac{1}{2}$ $XZ$-type systems on the hyper-cubic
lattice $\mathbb Z^d$ with a Gibbsian ground state, characterized
by the classical spin potential energy~$U_0(s_{\Lambda})$, in
which two lro's can occur for the spin operators $S^1$ and $S^3$
in a dimension greater than one. In such the systems there is
always the ferromagnetic lro for $S^1$ even for $d=1$ if a~simple
condition for $U_0$ holds. The Hamiltonians, determined as
symmetric matrices in the $2|\Lambda|$ dimensional complex Hilbert
space ${\mathbb C}^{2|\Lambda|}$ with the Euclidean scalar product
$(\cdot,\cdot)$, were given by
\begin{gather}
H_{\Lambda}=\sum_{ A\subset \Lambda,|A|>0}J_{A}P_{A}, \qquad
J_A\leq 0,\qquad
 P_A=S^1_{[A]}-e^{-\frac{\alpha}{2} W_A(S^3_{\Lambda})},\qquad S^1_{[A]}=\prod\limits_{x\in A}S^1_x,
\nonumber\\
W_A(S^3_{\Lambda})=U_0\big(S^{3A}_{\Lambda}\big)-U_0\big(S^3_{\Lambda}\big),\qquad
S^{3A}_{\Lambda}=\big(S^3_{\Lambda\backslash A}, -S^3_A\big).\label{eq1}
\end{gather}
$J_A$ are real numbers, $S^1_x$, $S^3_x$, $x\in {\mathbb Z}^d $
are the `unity' Pauli matrices, $\Lambda\subset \mathbb Z^d$ is a
hypercube with a finite cardinality $|\Lambda|$ (number of sites).
$S^1$ is diagonal and $S^3_{1,1}=-S^3_{2,2}=1$. $S^1$ has zero
diagonal elements and $S^1_{1,2}=S^1_{2,1}=1$. The matrices at
different sites commute. $S^l_{[A]}$ is the abbreviated notation
for the tensor product of the matrices~$S^l_x$, $x\in A$ and the
unity matrices~$I_x$, $x\in \Lambda\backslash A$.

The Gibbsian non-normalized state $\Psi_{\Lambda}$ is given by
\begin{gather}
\Psi_{\Lambda}=\sum\limits_{s_{\Lambda}}e^{-\frac{\alpha}{2}
U_0(s_{\Lambda})}\Psi^0_{\Lambda}(s_{\Lambda}),\qquad \alpha\in
\mathbb R^+,\label{eq2}
\end{gather}
where the summation is performed over $(\times
(-1,1))^{|\Lambda|}$,
\[
\Psi^0_{\Lambda}(s_{\Lambda})=\otimes_{x\in
\Lambda}\psi_0(s_x),\qquad \psi_0(1)=(1,0), \qquad
\psi_0(-1)=(0,1).
\]

These systems differ from the XZ spin $\frac{1}{2}$ systems, which
admit Gibbsian ground states considered in \cite{Kirkwood&Thomas}.
The potential energy of the associated classical Gibbsian system,
which generates the ground state, is found there in the form of a
perturbation expansion in a small parameter (an analog of
$\alpha$). The authors proved that there is the ferromagnetic lro
for $S^3$ in the ground state in some of their ferromagnetic
systems. Our proof of the $S^1$-lro is a simplified analog of
their proof. Uniqueness of a Gibbsian translation invariant ground
state is established in the thermodynamic limit for general $XZ$
models with a sufficiently strong magnetic field in~\cite{Matsui
1}.

In~\cite{Matsui 2} the classical Gibbsian states are identified
with ground states of quantum Potts models. The structure of the
considered Hamiltonians are close to the Hamiltonians of XZ spin
systems which are represented as a sum of a diagonal and
non-diagonal parts.

In this paper we consider the Hamiltonians
\begin{gather}
H_{\Lambda}=H_{0\Lambda}+V_{\Lambda},\qquad
H_{0\Lambda}=\sum\limits_{\mbox{\scriptsize $\begin{array}{c}
A,A'\subseteq \Lambda,\\ A\cap
A'=\varnothing\end{array}$}} \!\! \phi_{A,A'}S^1_{[A]}S^2_{[A']},\label{eq3}
\end{gather}
where $ \phi _{A,A'}$ are real valued coefficients, $S^2$ is the
second Pauli matrix with the zero diagonal elements such that
$S^2_{1,2}=-S^2_{2,1}=-i$ and $V_{\Lambda}$ depends on
$S^3_{\Lambda}$. We find the expression for $V_{\Lambda}$ which
guarantees that $\Psi_{\Lambda}$ given by~\eqref{eq2} is the eigen
(ground) state. This result is a generalization of our previous
result since with the help of~\eqref{eq2.1} we reduce our
Hamiltonian to the Hamiltonian~\eqref{eq1} with $J_A$ depending in
$S^3_{\Lambda}$. Our result is summarized in the following theorem

\medskip

\noindent
{\bf Theorem.} 
{\it Let $V_{\Lambda}$ be given by
\begin{gather}
V_{\Lambda}=-\sum\limits_{A\subseteq
\Lambda}J_{A}\big(S^3_A\big)e^{-\frac{\alpha}{2}
W_{A}(S^3_{\Lambda})},\qquad J_{A}\big(S^3_A\big)=\sum\limits_{A'\subseteq
A}(-i)^{|A'|}\phi_{A\backslash A',A'}S^3_{[A']}.\label{eq5}
\end{gather}
Then
\begin{enumerate}
\itemsep=0pt \item[\rm I.] $\Psi_{\Lambda}$ is an eigenfunction of
the Hamiltonian \eqref{eq3};

\item[\rm II.] $\Psi_{\Lambda}$ is its ground state if
$\phi_{A,A'}=0$ for odd $|A'|$ and $J_{A}\leq 0$;

\item[\rm III.] lro for $S^1$ occurs in the eigenstate
$\Psi_{\Lambda}$ if $\lim\limits_{\Lambda\rightarrow \mathbb
Z^d}W_{A}(s_{\Lambda})$ exists for $|A|=2$  and is uniformly
bounded. Moreover, $\langle S^1_{[A]}\rangle_{\Lambda}\geq a>0$,
where $a$ is a constant independent of $\Lambda$ if
$\lim\limits_{\Lambda\rightarrow \mathbb Z^d}W_{A}(s_{\Lambda})$
exists and is uniformly bounded.

\item[\rm IV.] lro occurs for $S^3$ in the eigenstate
$\Psi_{\Lambda}$ if lro occurs in the classical spin system with
the potential energy $U_0$.
\end{enumerate}}

If  $H_{0\Lambda}$ coincides with the Hamiltonian of the $XX$
Heisenberg model
\[
H_{0\Lambda}=\sum\limits_{x,y\in
\Lambda}\phi_{x,y}\big(S^1_xS^1_y+S^2_xS^2_y\big),
\]
then it can be shown without difficulty, utilizing the equality
$\big(S^3\big)^2=I$, that for the following choice
$U_0(s_{\Lambda})=\sum\limits_{x\in \Lambda}u_xs_x$ the matrix
$V_{\Lambda}$ in \eqref{eq5} is given by
\[
V_{\Lambda}=\sum\limits_{x,y\in
\Lambda}\phi_{x,y}\left[S^3_xS^3_y\cosh\alpha(u_x-u_y)-
(S^3_x-S^3_y)\sinh(\alpha (u_x-u_y))-
\cosh(\alpha(u_x-u_y))\right].
\]
If one puts $\phi_{x,y}=\phi_{x-y}$, $\phi_{x}=\phi_{|x|}=0$,
$|x|\not =1$, $\phi_{1}=J$, $q=e^{\alpha}$ and
\[
u_x=x^1+\cdots+x^d, \qquad x=\big(x^1,\ldots,x^d\big)
\]
then the following Hamiltonian of the $XXZ$ Heisenberg model is
derived
\begin{gather*}
H_{\Lambda}=J\sum\limits_{\langle x,y\rangle\in
\Lambda}\left[S_x^1S^1_y+S_x^2S^2_y+\frac{q+q^{-1}}{2}S^3_xS^3_y\right]\\
\phantom{H_{\Lambda}=}{}- 2J\sum\limits_{\langle x < y\rangle\in
\Lambda}\left[\frac{q-q^{-1}}{2}(S^3_x-S^3_y)+\frac{q+q^{-1}}{2}\right],
\end{gather*}
where the summations are performed over nearest neighbor pairs and
``$<$'' means lexicographic order. It is remarkable that the term
linear in $S^3_x$ contributes only on the boundary of $\Lambda$.
These Hamiltonians coincide with the Hamiltonians proposed
in~\cite{Alcaraz, Alcaraz&Salinas&Wrechinsky}. A reader may find
out that the Gibbsian ground states of these Hamiltonians are not
unique. Gibbsian ground states for a~partial case of our
Hamiltonians were considered in~\cite{Matsui 3}.

If $F_A$ depends on $S^3_A$, $S^1_A$ then its expectation value in
a state $\Psi_{\Lambda}$ is given by
\begin{gather*}
\langle F_A\rangle _{\Lambda}=(\Psi_{\Lambda},\Psi_{\Lambda})^{-1}
\big(\Psi_{\Lambda},F_A\big(S^1_A, S^3_{A}\big)\Psi_{\Lambda}\big),
\end{gather*}
where $(\cdot,\cdot)$ is the Euclidean scalar product on $\mathbb
C^{2|\Lambda|}$. Ferromagnetic lro for $S^l$ occurs if
\begin{gather}
\langle S^l_xS^l_y\rangle_{\Lambda}\geq a_l>0,\label{eq7}
\end{gather}
where the constants $a_l$ are independent of $\Lambda$. It implies
that the magnetization $M^l_{\Lambda}$ is an order parameter in
the thermodynamic limit since
\[
\langle (M^l_{\Lambda})^2\rangle_{\Lambda}\geq a_l>0,\qquad
M^l_{\Lambda}=|\Lambda|^{-1}\sum\limits_{x\in \Lambda}S^l_x.
\]
Besides, the inequality in the statement III of the Theorem for
$|A|=1$ implies that
\[
\langle M^1_{\Lambda}\rangle_{\Lambda}\geq a>0.
\]

It is well known that in the classical Ising model with a
ferromagnetic short-range potential energy $U_0$, generated by the
nearest neighbor bilinear pair potential, there is
the~ferromagnetic~lro. Hence our quantum systems for such $U_0$
admit two order parameters $M^l_{\Lambda}$, $l=1,3$ in
the~thermodynamic limit for sufficiently large $\alpha$ since for
such $U_0$ the condition in the statements III--IV of the Theorem
is true. For small values of $\alpha$ the magnetization in the
third direction vanishes for short-range pair interaction
potentials.

The last statement of the theorem follows without difficulty since
$S^3$ is a diagonal matrix and the ground state expectation value
$\langle F_A\rangle_{\Lambda}$ of a function $F_A$ depending on
$S^3_{A}$ equals the classical Gibbsian expectation value of the
same function depending on the classical spins $s_{A}$
correspon\-ding to the potential energy $U_0(s_{\Lambda})$ and the
``inverse temperature'' $\alpha$. The orthogonality of the basis
\[
(\Psi^0_{\Lambda}(s_{\Lambda}),\Psi^0_{\Lambda}(s'_{\Lambda}))=\prod\limits_{x\in
\Lambda}\delta_{s_x,s'_x},
\]
where $\delta_{s,s'}$ is the Kronecker symbol, has to be applied
for proving that.

The proofs of the second statement of the Theorem is based on the
proof that $H_{\Lambda}$ is positive definite. Its condition
implies that the semigroup generated by $H_{\Lambda}$ has positive
matrix elements. As a result the operator
\[
H^+_{\Lambda}=e^{\frac{\alpha}{2}U_0(S^3_{\Lambda})}H_{\Lambda}
e^{-\frac{\alpha}{2}U_0(S^3_{\Lambda})}.
\]
generates a Markovian process with a stationary state. We show
that it is symmetric and positive definite in the Euclidean scalar
product with the operator weight $e^{-\alpha U_0(S^3_{\Lambda})}$.

For sufficiently small $\alpha$ and a pair simple ferromagnetic
interaction with $\phi_{A,A'}=0$ for $|A'|\not =0$ lro at non-zero
low temperature occurs for $S^1$~\cite{Thomas&Yin}.

\section{Proof of Theorem}

It easy to check that
\begin{gather}
S^2=-iS^3S^1=iS^1S^3.\label{eq2.1}
\end{gather}
From this equality the following equalities are derived
\begin{gather*}
S_{[A']}^2=(-i)^{|A'|}S_{[A']}^3S_{[A']}^1,
\\
H_{0\Lambda}= \sum\limits_{\mbox{\scriptsize $\begin{array}{c}
A,A'\subseteq \Lambda,\\ A\cap A'=\varnothing\end{array}$}}\!\! (-i)^{
|A'|}\phi_{A,A'}S^1_{[A]}S^3_{[A']}S^1_{[A']}=
\sum\limits_{\mbox{\scriptsize $\begin{array}{c} A,A'\subseteq
\Lambda,\\ A\cap
A'=\varnothing\end{array}$}} \!\! (-i)^{|A'|}\phi_{A,A'}S^3_{[A']}S^1_{[A\cup
A']}.
\end{gather*}
As a result
\begin{gather}
H_{0\Lambda}=\sum\limits_{A\subseteq
\Lambda}J_{A}(S^3_{A})S^1_{[A]}.\label{eq2.2}
\end{gather}
\eqref{eq2.2} and \eqref{eq5} lead to the following expression for
the Hamiltonian \eqref{eq3}
\begin{gather}
H_{\Lambda}=\sum_{ A\subset \Lambda,\,
|A|>0}J_A(S^3_{A})P_A.\label{eq2.3}
\end{gather}

\noindent {\bf Proof of I.} $S^1$ flips spins:
\[
S^1_{[A]}\Psi^0_{\Lambda}(s_{\Lambda})=\Psi^0_{\Lambda}(s^A_{\Lambda})=
\Psi^0_{\Lambda}(s_{\Lambda\backslash A}, -s_A),\qquad
S^3_x\Psi^0_{\Lambda}(s_{\Lambda})=s_x\Psi^0_{\Lambda}(s_{\Lambda}).
\]
These identities lead to
\[
J_A\big(S^3_{A}\big)P_A\Psi_{\Lambda}=
\sum\limits_{s_{\Lambda}}\left(J_A(-s_{A})\Psi^0_{\Lambda}
(s_{\Lambda\backslash
A},-s_A)-J_A(s_{A})e^{-\frac{\alpha}{2}W_A(s_{\Lambda})}
\Psi^0_{\Lambda}(s_{\Lambda})\right)
e^{-\frac{\alpha}{2}U_0(s_{\Lambda})}.
\]

From the definition of $W_A$  and after changing signs of the spin
variables $s_A$ in the first term in the sum it follows that
\begin{gather*}
J_A\big(S^3_{A}\big)P_A\Psi_{\Lambda}=\sum\limits_{s_{\Lambda}}\left[J_A(-s_{A})
\Psi^0_{\Lambda} (s_{\Lambda\backslash
A},-s_A)e^{-\frac{\alpha}{2}U_0(s_{\Lambda})}
-J_A(s_{A})\Psi^0_{\Lambda}(s_{\Lambda})e^{-\frac{\alpha}{2}U_0(s^A_{\Lambda})}\right]
\\
\phantom{J_A(S^3_{A})P_A\Psi_{\Lambda}}{}
 =\sum\limits_{s_{\Lambda}}J_A(s_{A})\left(e^{-\frac{\alpha}{2}U_0(s^A_{\Lambda})}
-e^{-\frac{\alpha}{2}U_0(s^A_{\Lambda})}\right)\Psi^0_{\Lambda}(s_{\Lambda})=0.
\end{gather*}
That is, every term in the sum for $H_{\Lambda}\Psi_{\Lambda}$
in~\eqref{eq2.3} is equal to zero. This proves the statement.

\medskip

\noindent {\bf Proof of II.} It is necessary to prove that the
Hamiltonian is positive-definite. For that purpose we'll use the
operator $H^+_{\Lambda}$ introduced in the introduction. It is not
difficult to check on the basis~$\Psi^0_{\Lambda}$ that
\begin{gather*}
H^+_{\Lambda}=\sum\limits_{A\subseteq \Lambda}J_A\big(S^3_A\big)
e^{-\frac{\alpha}{2}W_A(S^3_{\Lambda})}\big(S^1_{[A]}-I\big),
\end{gather*}
where $I$ is the unity operator. If
\[
F=\sum\limits_{s_{\Lambda}}F(s_{\Lambda})\Psi^0_{\Lambda}(s_{\Lambda}),
\qquad
H^+_{\Lambda}F=\sum\limits_{s_{\Lambda}}(H^+_{\Lambda}F)(s_{\Lambda})
\Psi^0_{\Lambda}(s_{\Lambda})
\]
then, taking into account that $J_A(s_A)$ is en even function in
$s_x$, we obtain
\[
(H^+_{\Lambda}F)(s_{\Lambda})=-\sum\limits_{A\subseteq
\Lambda}J_A(s_{A})e^{-\frac{\alpha}{2}
W_A(s_{\Lambda})}\big(F(s_{\Lambda})-F\big(s^A_{\Lambda}\big)\big).
\]
$H^+_{\Lambda}$ is symmetric with respect to the new scalar
product
\[
(F, F')_{U_0}=\big(e^{-\alpha U_0(S^3_{\Lambda})}F, F'\big).
\]
The check is given by
\begin{gather*}
(H^+_{\Lambda}F, F')_{U_0}=\big(e^{-\alpha
U_0(S^3_{\Lambda})}H^+_{\Lambda}F, F'\big)= \sum\limits_{A\subseteq
\Lambda}\big(J_A(S^3_{A})
e^{-\frac{\alpha}{2}[U_0(S^3_{\Lambda})+U_0(S^{3A}_{\Lambda})]}(S^1_{[A]}-
I)F,F'\big)
\\
\phantom{(H^+_{\Lambda}F, F')_{U_0}}{} =\sum\limits_{A\subseteq
\Lambda}\big(J_A(S^3_{A})
e^{-\frac{\alpha}{2}[U_0(S^3_{\Lambda})+U_0(S^{3A}_{\Lambda})]}F,(S^1_{[A]}-
I)F'\big)=(F, H^+_{\Lambda}F')_{U_0}.
\end{gather*}
Here we used the equalities
\begin{gather*}
e^{-\frac{\alpha}{2}U_0(S^3_{\Lambda})}S^1_{[A]}=
S^1_{[A]}e^{-\frac{\alpha}{2}U_0(S^{3A}_{\Lambda})}, \qquad
e^{-\frac{\alpha}{2}U_0(S^{3A}_{\Lambda})}S^1_{[A]}=
S^1_{[A]}e^{-\frac{\alpha}{2}U_0(S^3_{\Lambda})}
\end{gather*}
and the fact that $S^1_{[A]}$ commutes with $J_{A}(S^3_{A})$. From
the definitions it follows that
\begin{gather}
(H^+_{\Lambda}F,F')_{U_0}=
\big(H_{\Lambda}e^{-\frac{\alpha}{2}U_0(S^3_{\Lambda})}F,
e^{-\frac{\alpha}{2}U_0(S^3_{\Lambda})}F'\big).\label{eq2.6}
\end{gather}
$H_{\Lambda}^+$ is positive definite. This is a consequence of the
relations
\begin{gather}
(H^+_{\Lambda}F,F)_{U_0}=-\sum\limits_{A\subseteq
\Lambda}\sum\limits_{s_{\Lambda}}J_A(s_{A})e^{-\frac{\alpha}{2}
[U_0(s_{\Lambda})+U_0(s^A_{\Lambda})]}\big(F(s_{\Lambda})-F\big(s^A_{\Lambda}\big)\big)
F(s_{\Lambda})\nonumber\\
\phantom{(H^+_{\Lambda}F,F)_{U_0}}{}=-\frac{1}{2}\sum\limits_{A\subseteq
\Lambda}\sum\limits_{s_{\Lambda}}J_A(s_{A})e^{-\frac{\alpha}{2}
[U_0(s_{\Lambda})+U_0(s^A_{\Lambda})]}\big(F(s_{\Lambda})-
F\big(s^A_{\Lambda}\big)\big)^2\geq 0.\label{eq2.7}
\end{gather}
Here we took into account that the exponential weight in the sum
is invariant under changing signs of spin variables $s_A$. From
\eqref{eq2.6}, \eqref{eq2.7} it  follows that $H_{\Lambda}$ is,
also, positive definite. Statement is proved.

\medskip

\noindent {\bf Proof of III.}  We have to prove \eqref{eq7} for
$l=1$. From orthogonality of the basis it follows that
\[
Z_{\Lambda}=(\Psi_{\Lambda},\Psi_{\Lambda})=\sum\limits_{s_{\Lambda}}
e^{-\alpha U_0(s_{\Lambda})}
\]
and
\[
\langle S^1_{[A]}\rangle_{\Lambda}=Z^{-1}_{\Lambda}
\sum\limits_{s_{\Lambda}}e^{-\alpha
U_0(s_{\Lambda})}e^{-\frac{\alpha}{2} W_{A}(s_{\Lambda})}\geq
\inf\limits_{s_{\Lambda},A}e^{-\frac{\alpha}{2}
W_{A}(s_{\Lambda})}\geq
e^{-\frac{\alpha}{2}\max\limits_{s_{\Lambda}, A}
|W_{A}(s_{\Lambda})|}.
\]
This proves the statement.

\section{Discussion}

The proposed perturbations $V_{\Lambda}$ of the initial
Hamiltonian $H_{0\Lambda}$ seem complicated. But it is not always
so. If the initial Hamiltonian coincides with the Hamiltonian of
the $XX$ Heisenberg model, then in some cases, mentioned in the
introduction, the perturbation is very simple and produces an
anisotropic quadratic in $S^3_x$, $x\in \Lambda$ term and a
boundary term linear in $S^3_x$, $x\in \Lambda$ only. A reduction
of $V_{\Lambda}$ to a simpler form  for a quadratic in $S^3_x$,
$x\in \Lambda$ function $U_0$ can be found in
\cite{Dorlas&Skrypnik}. There is an interesting problem to find
out all the cases of the initial Hamiltonians, commuting with the
total spin in the third direction, $\phi_{A,A'}$ and $U_0$ leading
to simple anisotropic generalized $XXZ$ models.

\LastPageEnding


\begin{thebibliography}{99}
\footnotesize

\bibitem{Dorlas&Skrypnik} Dorlas T., Skrypnik W., Two order parameters in quantum XZ spin
midels with Gibbsian ground states, {\it J.~Phys.~A: Math. Gen.},
2004, V.37, 6623--6632.

\bibitem{Kirkwood&Thomas} Kirkwood J., Thomas L., Expansions and phase transitions for the
ground state of quantum Ising lattice systems, {\it Comm. Math.
Phys.}, 1983, V.88, 569--580.

\bibitem{Matsui 1}Matsui T., A link between quantum and classical Potts models,
{\it J.~Statist. Phys.}, 1990, V.59, 781--798.

\bibitem{Matsui 2} Matsui T., Uniqueness of translation invariant ground
state in quantum spin systems, {\it Comm. Math. Phys.}, 1990,
V.126, 453--467.

\bibitem{Alcaraz}Alcaraz F., Exact steady states of asymmetric diffusion and
two-species annihilation with back reaction from the ground state
of quantum spin model, {\it Internat. J. Modern Phys.}, 1994,
V.25--26, 3449--3461.

\bibitem{Alcaraz&Salinas&Wrechinsky}Alcaraz F., Salinas S., Wrechinsky W., Anisotropic quantum
domains, {\it Phys. Rev. Lett.}, 1995, V.5, 930--933.

\bibitem{Matsui 3} Matsui T., On ground state degeneracy of $Z_2$
symmetric quantum spin models, {\it Publ. Res. Inst. Math. Sci.},
1991, V.27, 658--679.

\bibitem{Thomas&Yin} Thomas~L., Yin Z., Low temperature
expansions for the Gibbs states of quantum Ising lattice systems,
{\it J.~Math. Phys.}, 1984, V.10, 3128--3134.

\end{thebibliography}
\end{document}